\documentclass[aps,prl,twocolumn,amsmath,amssymb,superscriptaddress,floatfix]{revtex4}

\usepackage{graphicx}

\begin{document}

\title{\Large\textbf\boldmath 
Coupling of Magnetic Orders in La$_{2}$CuO$_{4+x}$}

\author{Vyacheslav~G. Storchak}
\email{mussr@triumf.ca}
\affiliation{National Research Center ``Kurchatov Institute'', 
 Kurchatov~Sq. 1, Moscow 123182, Russia}

\author{Jess~H. Brewer}
\affiliation{Department of Physics and Astronomy, 
 University of British Columbia, 
 Vancouver, BC V6T 1Z1, Canada}

\author{Dmitry~G. Eshchenko}
\affiliation{Bruker BioSpin AG, 
 Industriestrasse 26, 8117 F\"allanden, Switzerland}

\author{Patrick~W. Mengyan}
\affiliation{Department of Physics, Texas Tech University, 
 Lubbock, TX 79409-1051, US}

\author{Oleg~E. Parfenov}
\affiliation{National Research Center ``Kurchatov Institute'', 
 Kurchatov~Sq. 1, Moscow 123182, Russia}

\author{Andrey~M. Tokmachev}
\affiliation{National Research Center ``Kurchatov Institute'', 
 Kurchatov~Sq. 1, Moscow 123182, Russia}

\author{Pinder Dosanjh}
\affiliation{Department of Physics and Astronomy, 
 University of British Columbia, 
 Vancouver, BC V6T 1Z1, Canada}


\date{08 June 2016}

\begin{abstract}

High transverse magnetic field and zero field muon spin rotation and 
relaxation measurements 
have been carried out in a lightly oxygen-doped high-$T_{c}$ parent compound 
La$_2$CuO$_{4}$ in a temperature range from 2~K to 300~K. As in the 
stoichiometric compound, muon spin rotation spectra reveal, along with the 
antiferromagnetic local field, the presence of an additional source of magnetic 
field at the muon. The results indicate that this second 
magnetic order is driven by the antiferromagnetism at low temperature but 
the two magnetic orders  
decouple at higher temperature. The ability of $\mu^+$SR to detect this 
additional magnetism deteriorates with doping, thus rendering the technique 
impotent to reveal time-reversal symmetry breaking in superconductors.

\end{abstract}

\maketitle

Superconducting (SC) cuprates exhibit a pseudogap (PG) state with anomalous 
transport, magnetic, optical and thermodynamic properties \cite{Timusk1999,
Norman2005}. This enigmatic state is believed to hold the key to the mechanism 
of high-temperature SC (HTSC) but its nature is still a major unsolved problem in condensed matter 
physics. Some theories suggest that the PG is a disordered precursor to the SC 
phase lacking phase coherence among preformed pairs \cite{Emery1995,Lee2006}. 
However, mounting experimental evidence associates the PG with a broken symmetry 
state accompanied by onset of charge density wave, nematic or magnetic orders 
\cite{Fauque2006,Xia2008,Daou2010,Neto2014,Comin2015,Wu2015}.

In particular, the magnetic order causes time-reversal symmetry breaking 
(TRSB). Angle-resolved photoemission spectra of 
Bi$_2$Sr$_2$CaCu$_2$O$_{8+x}$ demonstrate spontaneous dichroism, an 
indication of a TRSB ordered state \cite{Kaminski2002}. Polarized neutron 
scattering (NS) experiments in YBa$_2$Cu$_{3}$O$_{6+x}$ \cite{Fauque2006}, 
HgBa$_2$CuO$_{4+x}$ \cite{Li2008}, La$_{2-x}$Sr$_{x}$CuO$_{4}$ 
\cite{Baledent2010} and Bi$_2$Sr$_2$CaCu$_2$O$_{8+x}$ \cite{Almeida2012} 
reveal an intra-unit-cell magnetic order. Its onset coincides with the known PG 
boundary $T^{*}$ \cite{Shekhter2013}. The electronic state identified in those 
experiments is qualitatively consistent with the model of orbital current loops 
for the PG state \cite{Varma2006,Varma2014}, gaining further support from weak 
magnetic excitations detected in cuprates \cite{Li2010,Li2012}. The puzzling 
part is the observation of large in-plane components of magnetic moments. 
Possible explanations include a spin component due to spin-orbit coupling 
\cite{Aji2007}, a quantum superposition of loop currents \cite{He2012}, 
or a contribution from apical oxygen atoms \cite{Weber2009}. A recent study 
\cite{Mangin2015} suggests that $T^{*}$ corresponds only to the onset of the 
in-plane component.

Additional evidence for broken symmetry in the PG region comes from 
high-precision polar Kerr effect (PKE) measurements of cuprates \cite{Xia2008,
He2011,Karapetyan2014}. The effect signals TRSB but demonstrates unusual 
characteristics ascribed to a chiral order \cite{Karapetyan2014}. The relation 
between the PKE and NS observations is unclear: 
The characteristic PKE-detected moments are tiny, 
about 4 orders of magnitude smaller than those revealed in the NS experiments. 
The onset of PKE occurs at a temperature which is noticeably lower than $T^{*}$ 
but is close to that of charge ordering, prompting proposals in which 
fluctuating charge- \cite{Wang2014} or pair-density-wave \cite{Agterberg2015} 
states induce spontaneous currents with broken mirror symmetries.

Somewhat surprisingly, the magnetic order eludes detection with 
local magnetic probe techniques, 
thus arousing legitimate doubts on its intrinsic and universal 
nature. The failure of nuclear magnetic resonance (NMR) \cite{Mounce2013,
Wu2015} is certainly conspicuous --- upper bounds on static fields at oxygen 
sites are two orders of magnitude smaller than estimates for the current-loop 
order \cite{Wu2015}. Similarly unsuccessful is the search for orbital currents 
with the Zeeman-perturbed nuclear quadrupole technique \cite{Strassle2011}. It 
can be explained by a fluctuating character of the magnetic order, possibly 
induced by defects \cite{Varma2014}, and a large difference in the 
characteristic correlation times accessible by NMR and NS: 
the fluctuations may be too fast for NMR causing dynamical narrowing 
but fall within the time window of the NS technique.

In terms of the characteristic correlation times, muon spin relaxation  
($\mu^+$SR) techniques bridge the gap between NMR and NS, thus seeding 
expectations that the magnetic order observed with NS may leave pronounced 
fingerprints in $\mu^+$SR spectra. However, $\mu^+$SR experiments in 
highly doped HTSC cuprates \cite{MacDougall2008,Sonier2009} have not detected the 
expected magnetic order. Among the explanations put forward are again the 
defect-driven fluctuating character of the order \cite{Varma2014} but also 
screening of the charge density in the vicinity of the muon 
\cite{Shekhter2008}. Both problems are absent in the insulating parent 
compounds while the ordering may well be present should it be an intimate 
feature of chemical bonding in CuO$_2$ planes. Indeed, orbital currents have 
been observed in the antiferromagnetic (AFM) phase of insulator CuO \cite{Scagnoli2011}. 
Following this strategy we recently carried out $\mu^+$SR measurements on 
single crystals of another parent compound, stoichiometric La$_2$CuO$_{4}$ 
\cite{Storchak2015}. The transverse-field measurements show characteristic 
splittings in the spectra indicating the presence of a source of magnetic field 
\textit{additional to AFM}. The estimated magnitude and tilting of the 
local moments are found to match those detected in the NS experiments 
\cite{Storchak2015}. The main interest, however, concerns doped cuprates
which exhibit or approach the PG state.  

In this Letter we present the results of $\mu^+$SR studies of doped 
samples, namely La$_2$CuO$_{4+x}$. We follow the evolution of the spectra 
with doping ($x$), temperature and external magnetic field and demonstrate that the 
local muon probe does detect a magnetic order distinct from AFM in doped 
cuprates. The results also set limitations on the applicability of 
$\mu^+$SR spectroscopy to such problems.

The current $\mu^+$SR studies employ stoichiometric 
La$_2$CuO$_{4}$ and oxygen-doped La$_2$CuO$_{4+x}$ with $x$=0.0075 and 
$x$=0.0085. Higher doping sets an insurmountable hurdle 
for the $\mu^+$SR technique (see below). 
Single crystals of La$_2$CuO$_{4+x}$ are grown from CuO flux. The 
crystal orientation, lattice parameters, and mosaicity (not exceeding 
0.05$^\circ $ along the $\hat{c}$ axis) are determined with x-ray 
diffractometry. To produce stoichiometric samples, the surplus oxygen is removed 
by annealing in vacuum for 168~h at 700~$^\circ $C. The samples with $x$=0.0075 
come from annealing in air for 6~h at 900~$^\circ $C, while $x$=0.0085 is 
reached by similar annealing in oxygen ($p$=1~atm). The oxygen concentration is 
determined from the lattice parameter $c$ of orthorhombic La$_2$CuO$_{4+x}$ 
using Vegard's law \cite{Nikonov2000}. Time-differential $\mu^+$SR 
experiments, employing 100\% spin-polarized positive muons, were carried out on 
the M15 surface muon channel at TRIUMF using the {\it HiTime\/} spectrometer.

\begin{figure}
 \includegraphics[width=1.0\columnwidth]{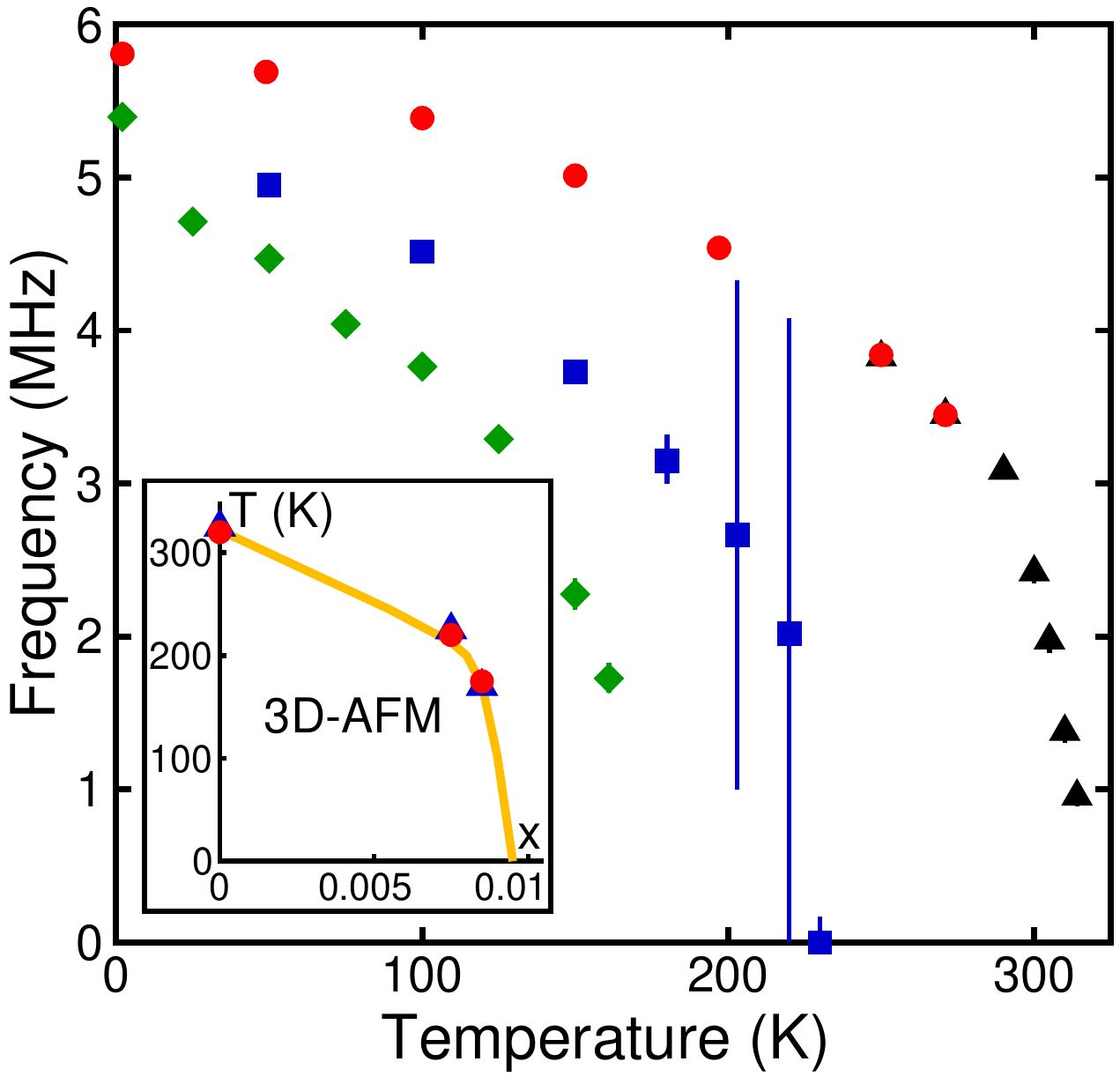}
 \caption{\label{zftrend}
  Temperature dependence of the muon spin precession frequency in ZF in 
  La$_2$CuO$_{4}$ (red circles for 2-component fit and black triangles for 
  1-component fit), La$_2$CuO$_{4.0075}$ (blue squares) and 
  La$_2$CuO$_{4.0085}$ (green diamonds). Inset: N\'eel temperatures 
  determined for the samples with SQUID (red circles) and ZF-$\mu^+$SR (blue 
  triangles) plotted against the boundary line for the 3D-AFM state of 
  La$_2$CuO$_{4+x}$.}
\end{figure}

The AFM behavior of the samples is well characterized by zero field (ZF) 
$\mu^+$SR spectroscopy. Positive muons, 
being a local microscopic magnetic probe, 
have proved to be remarkably sensitive to any kind of magnetic order. In 
La$_2$CuO$_{4}$ the muon stopping site is at a bonding distance from an 
apical oxygen on the plane bisecting an O-Cu-O angle of the copper-oxygen 
plaquette \cite{Storchak2015}. Figure \ref{zftrend} demonstrates the 
temperature dependence of the ZF muon spin precession frequency in all three 
samples. In the case of stoichiometric La$_2$CuO$_{4}$, ZF-$\mu^+$SR 
spectra at low temperature contain an additional small-amplitude component 
associated with a second muon site \cite{Storchak2015}. This signal disappears 
below the background at higher temperature in La$_2$CuO$_{4}$ and is not 
detected at all in the doped samples. The additional component is not shown 
in Figure \ref{zftrend} but necessitates a 2-component fit of ZF-$\mu^+$SR 
spectra of La$_2$CuO$_{4}$ at low temperature. The N\'eel temperatures  
determined from the temperature dependence of muon frequencies are 
325$\pm $5~K, 225$\pm $5~K and 170$\pm $5~K for La$_2$CuO$_{4}$, 
La$_2$CuO$_{4.0075}$ and La$_2$CuO$_{4.0085}$, respectively. These values are 
in full agreement with magnetization measurements: the inset of Figure 
\ref{zftrend} shows the N\'eel temperatures of all 3 samples determined with 
SQUID and ZF-$\mu^+$SR superimposed on the phase diagram of 
La$_2$CuO$_{4+x}$ \cite{Nikonov2000}.

\begin{figure}
 \includegraphics[width=0.929\columnwidth]{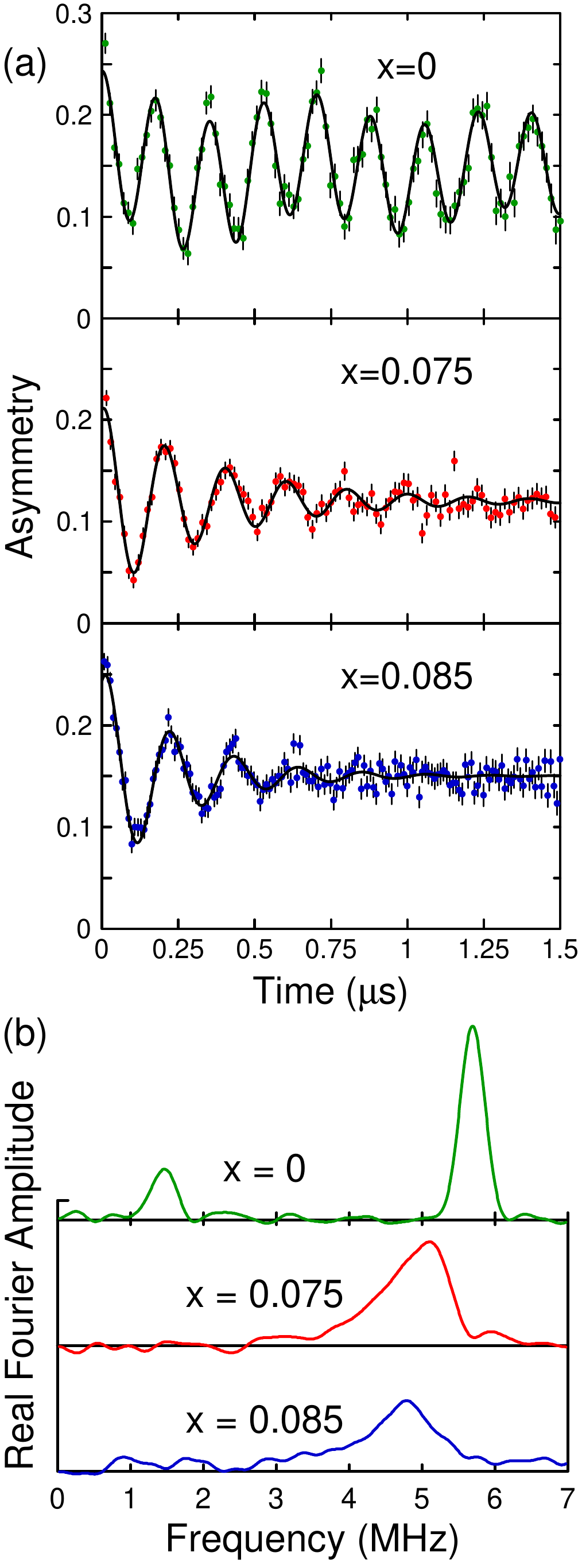}
 \caption{\label{zfbroaden}
  ($a$) Time-domain spectra of muon spin precession in ZF at 50~K in La$_2$CuO$_{4}$ 
  (green line), La$_2$CuO$_{4.0075}$ (red line), and La$_2$CuO$_{4.0085}$ 
  (blue line); ($b$) Fourier transforms of the same spectra.}
\end{figure}

The difference between the samples is not limited to signal frequencies and 
their temperature dependences --- the envelope of spectra also changes. Figure 
\ref{zfbroaden} shows the evolution of ZF-$\mu^+$SR spectra at 50~K with 
oxygen doping, in both time and frequency domains. One can see that even small 
doping results in significant 
broadening of the spectra. Probably inhomogeneities in the oxygen distribution cause 
magnetic field inhomogeneities, increasing the linewidth of the $\mu^+$SR signal.  
Such $\mu^+$SR line broadening may prevent detection of magnetic order, 
especially in heavily doped cuprates.

Like the NMR studies, broad ZF-$\mu^+$SR spectra do not reveal any additional 
magnetism (AM). However, in order to reconcile the experimental facts 
accumulated so far, one has to appreciate pecularities of the techniques. 
Indeed, comparatively long characteristic times may excuse NMR, but the 
$\mu^+$SR technique should be able to do the job when the sample is 
close to be insulating so that the charge screening effect does not apply. In 
fact, $\mu^+$SR is expected to be better suited for the task than neutrons 
as it ``measures'' in real space and ``sees'' roughly only the first 
coordination sphere around the muon, while NS is essentially a $k$-space 
technique requiring a substantial correlation length 
for the neutron to be effective as a magnetic probe.

\begin{figure}
 \includegraphics[width=1.0\columnwidth]{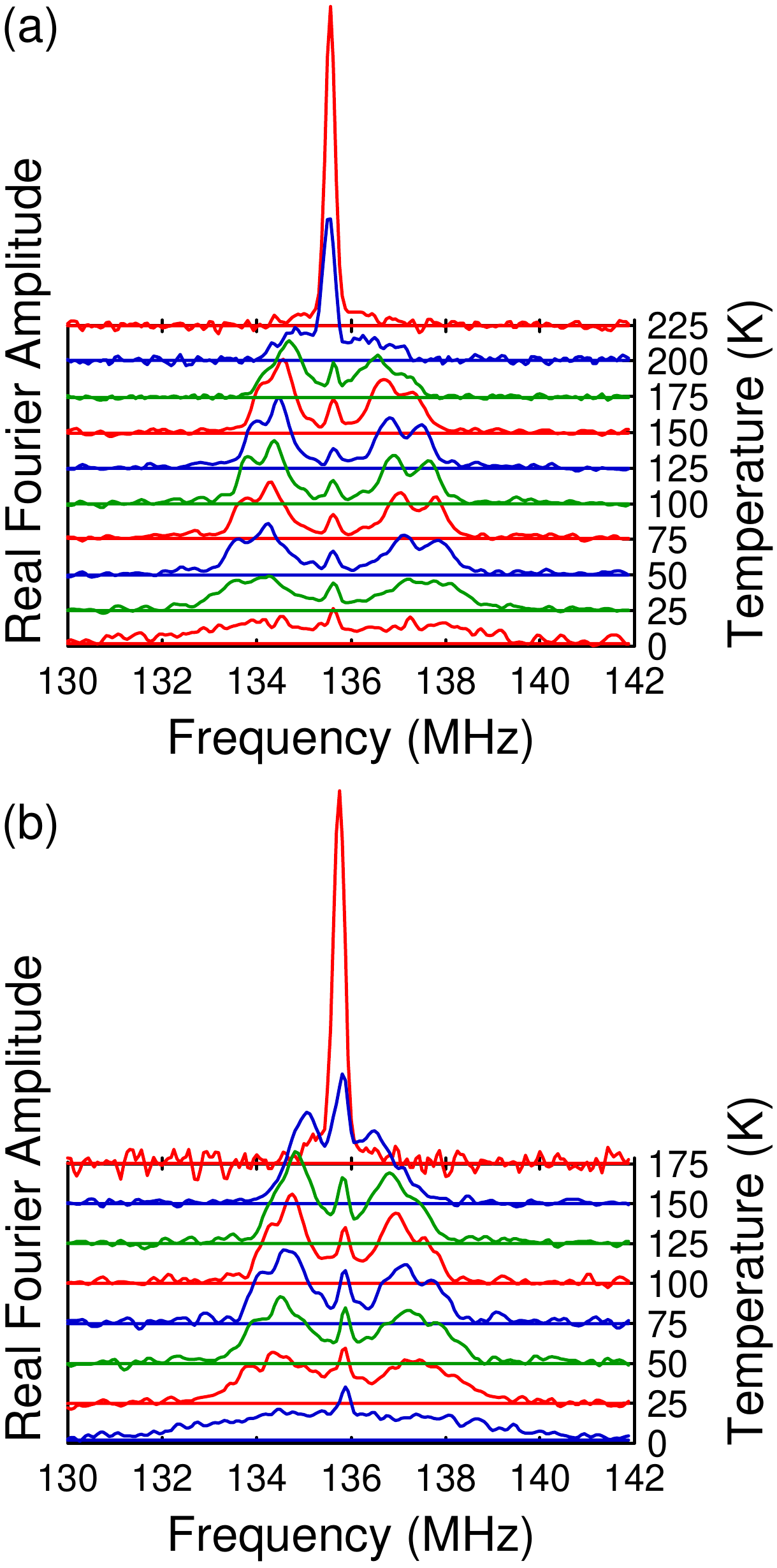}
 \caption{\label{panels}
  Fourier transforms of the muon spin precession signal in ($a$) 
  La$_2$CuO$_{4.0075}$ and ($b$) La$_2$CuO$_{4.0085}$ in an external magnetic 
  field of 1~T directed along the $\hat{c}$ axis of the crystal at different 
  temperatures.}
\end{figure}

For detection of AM we resort to transverse field (TF) $\mu^+$SR studies, 
which are often helpful in revealing differences in local magnetic fields 
that are hidden from ZF-$\mu^+$SR spectra. Figure \ref{panels} presents 
$\mu^+$SR spectra for the doped samples in a transverse magnetic field of 
1~T at different temperatures. The corresponding spectra for the stoichiometric 
La$_2$CuO$_{4}$ are given in Ref. \cite{Storchak2015}. The central line at 
about 135.6 MHz comes from muons that miss the sample and stop in a nonmagnetic 
environment. In the case of AFM there should be only two signals besides the 
central one. Additional peaks indicate the presence of AM. Namely, each of the two 
AFM signals on both sides of the central line is further split into two. 
In fact, the spectra are those expected for a combination of the AFM order 
and the AM detected by NS in highly doped cuprates. 
As in the ZF spectra, doping leads to broadening of the signals 
and the extra (additional to AFM) splitting becomes smeared. 
The study of signal splittings in La$_2$CuO$_{4}$ for different 
directions of the external magnetic field \cite{Storchak2015} allowed us to 
determine the magnetic field vector at the muon. This information is 
sufficient to rule out some of the proposed models for AM in cuprates 
\cite{Storchak2015}. However, all those conclusions are valid only if the 
splitting is indeed caused by AM. One can imagine that the splitting pattern 
comes not from an independent magnetism but from the same AFM moments acting 
upon muons in two different structural positions --- arising, for example, from 
two different tiltings of CuO$_{6}$ octahedra \cite{Reehuis2006}. The absence 
of any signal splittings above the N\'eel temperature certainly adds 
credibility to this alternative. Although the observed splitting is too 
large for such an explanation \cite{Storchak2015}, further studies 
are necessary to exclude such a possibility.

The hypotheses of AM \textit{vs.} two inequivalent sites can be verified by combined analysis 
of the temperature dependence of the splittings, 
especially in the vicinity of the N\'eel temperature. TF-$\mu^+$SR 
spectra allow determination of the component of the local magnetic field on 
the muon $B_{\parallel }$ along the external magnetic field $B_{\rm ext}$. The 
amplitude of a TF-$\mu^+$SR signal is 
$B=\sqrt{(B_{\parallel }+B_{\rm ext}) ^{2}+B_{\perp }^{2}}$, which means that the 
component $B_{\parallel }$ can be evaluated as $(B^{2}-B_{0}^{2}-B_{\rm ext}^{2})/
2B_{\rm ext}$, where $B_{0}$ is the modulus of the local magnetic field as 
given by ZF-$\mu^+$SR. To characterize the magnetic structure we 
determine the 4 signals coming from AFM and (supposedly) AM by fitting 
TF-$\mu^+$SR spectra in the time domain. Then, 4 magnetic field projections 
$B_{\parallel }^{I}$, $B_{\parallel }^{II}$, $B_{\parallel }^{III}$ and 
$B_{\parallel }^{IV}$ (in ascending order) are computed and the average 
splittings associated with the AFM and AM magnetic orders are defined as 
$(B_{\parallel }^{IV}+B_{\parallel }^{III}-B_{\parallel }^{II}-B_{\parallel }^{I})/2$ and 
$(B_{\parallel }^{IV}-B_{\parallel }^{III}+B_{\parallel }^{II}-B_{\parallel }^{I})/2$, 
respectively.

\begin{figure}
 \includegraphics[width=1.0\columnwidth]{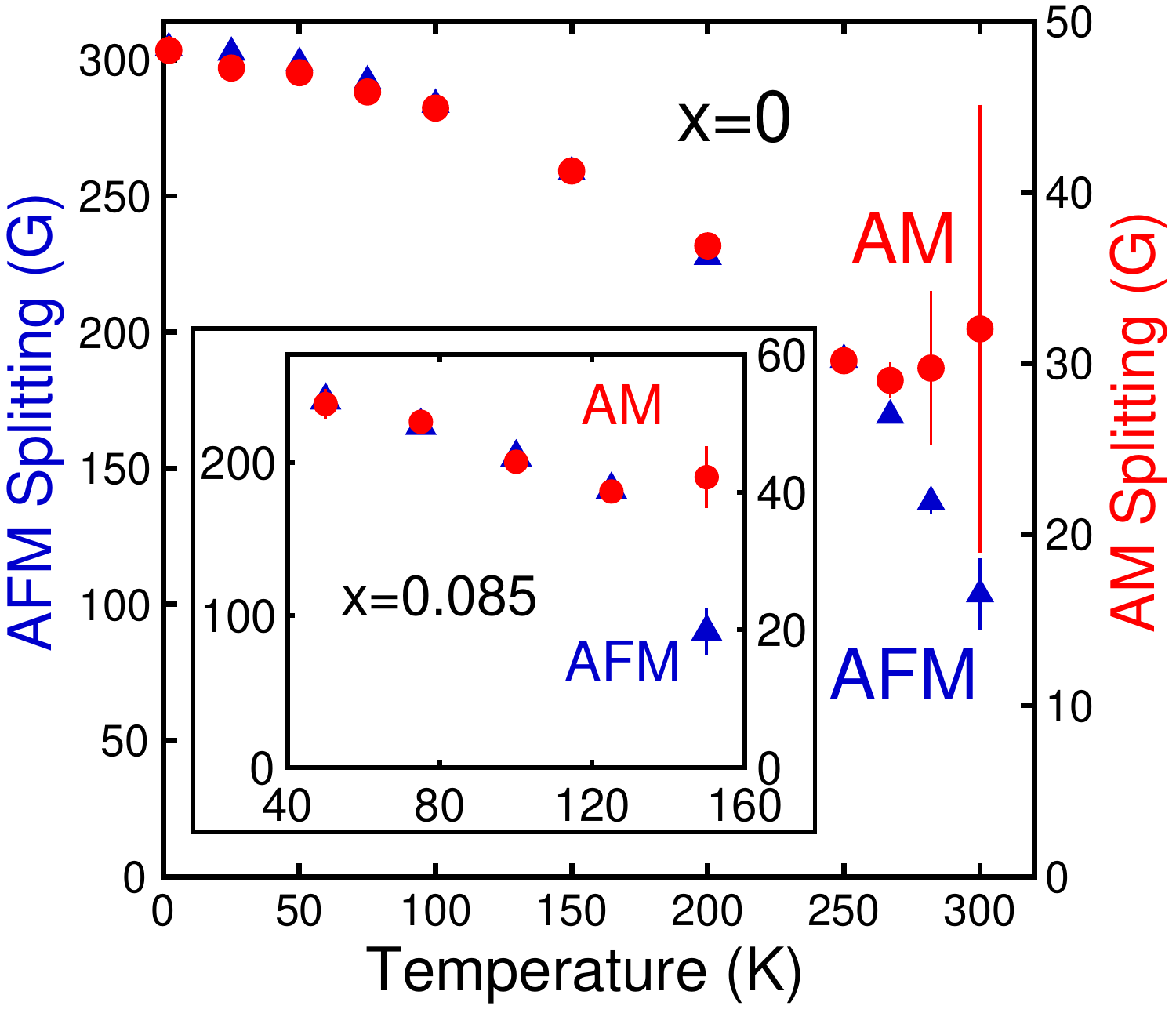}
 \caption{\label{orders}
  Temperature dependence of the splittings in TF-$\mu^+$SR spectra of 
  La$_2$CuO$_{4}$ caused by AFM (blue triangles) and AM (red circles) orders. 
  Inset: the same dependences for La$_2$CuO$_{4.0085}$.}
\end{figure}

Figure \ref{orders} shows the temperature dependence of the two splittings for 
stoichiometric La$_2$CuO$_{4}$. The behavior is quite peculiar. One can 
distinguish two regions: below 250~K the splittings are proportional to each 
other but above 250~K there is a sharp divergence of the trends. The behavior 
within the higher temperature region probably rules out the hypothesis of two 
structural muon positions and a single AFM order. Similar temperature 
dependences are observed for the doped samples --- the inset of Figure 
\ref{orders} shows it for La$_2$CuO$_{4.0085}$. The coupling of the two 
magnetic orders seems to be largely determined by the strength of the AFM 
order: at lower temperature the AM is driven by AFM but in the region close to the 
N\'eel temperature, the AFM order is rapidly dying out and the AM order decouples 
from the AFM order. The AM splitting even {\sl increases\/} when the N\'eel temperature 
is approached. However, the AM order is not observed above the N\'eel 
temperature. This does not necessarily mean that it is absent, 
only that $\mu^+$SR techniques are not capable to detect any AM. 
It is reasonable to suppose that AFM affects the fluctuation time 
of the AM order: without the AFM order the 
characteristic fluctuation times of AM are too small 
for this magnetism to be detected directly with 
$\mu^+$SR (in contrast to neutrons). Regrettably, it also means that 
$\mu^+$SR techniques stand no chance in finding this AM in heavily doped cuprates.

In summary, we studied local magnetic fields in lightly oxygen-doped as well as 
stoichiometric La$_2$CuO$_{4}$ with ZF- and TF-$\mu^+$SR spectroscopy. 
Both techniques demonstrate that doping leads to an increase in 
magnetic field inhomogeneity that hinders detection of 
magnetic ordering. TF-$\mu^+$SR experiments show a 
characteristic pattern based on 5 signals: a central line from muons that 
missed the sample and 4 signals corresponding to AFM superimposed with some 
additional magnetic order. 
The temperature dependence of the spectral lines reveals 
that the two magnetic orders are strongly coupled at low temperature. However, 
when the AFM order is weakened at higher temperature, the second magnetic order 
gains strength. Thus, we assert the existence of an additional 
magnetic order stemming from the AFM phase. 
The result is especially important since recent Hall coefficient 
measurements establish that the pseudogap in cuprates is a separate phenomenon 
from the charge order but strongly linked with the AFM Mott insulator 
\cite{Badoux2016}. Our results also explain the failure of previous attempts to 
detect the magnetic order in doped cuprates with $\mu^+$SR: the technique is 
capable of its detection only when the doping level is relatively small.

This work was partially supported by the Kurchatov Institute, 
the Russian Science Foundation (Grant 14-19-00662), 
the Russian Foundation for Basic Research (Grants 16-07-00204 and 16-29-03027), 
NSERC of Canada 
and the U.S. DOE, Basic Energy Sciences (Grant DE-SC0001769).


\begin{thebibliography}{99}

\bibitem{Timusk1999}
T. Timusk and B. Statt, 
Rep Prog. Phys. \textbf{62}, 61 (1999).

\bibitem{Norman2005}
M.~R. Norman, D. Pines, and C. Kallin, 
Adv. Phys. \textbf{54}, 715 (2005).

\bibitem{Emery1995}
V.~J. Emery and S.~A. Kivelson, 
Nature \textbf{374}, 434 (1995).

\bibitem{Lee2006}
P.~A. Lee, N. Nagaosa, and X.-G. Wen, 
Rev. Mod. Phys. \textbf{78}, 17 (2006).

\bibitem{Fauque2006}
B. Fauqu\'e, Y. Sidis, V. Hinkov, S. Pailh\`es, C.~T. Lin, X. Chaud, and P. Bourges, 
Phys. Rev. Lett. \textbf{96}, 197001 (2006).

\bibitem{Xia2008}
J. Xia \textit{et al.}, 
Phys. Rev. Lett. \textbf{100}, 127002 (2008).

\bibitem{Daou2010}
R. Daou \textit{et al.}, 
Nature \textbf{463}, 519 (2010).

\bibitem{Neto2014}
E.~H. da~Silva~Neto \textit{et al.}, 
Science \textbf{343}, 393 (2014).

\bibitem{Comin2015}
R. Comin \textit{et al.}, 
Nature Mat. \textbf{14}, 796 (2015).

\bibitem{Wu2015}
T. Wu, H. Mayaffre, S. Kr\"amer, M. Horvati\'c, C. Berthier, W.~N. Hardy, R. Liang, D.~A. Bonn, and M.-H. Julien, 
Nature Comm. \textbf{6}, 6438 (2015).

\bibitem{Kaminski2002}
A. Kaminski \textit{et al.}, 
Nature \textbf{416}, 610 (2002).

\bibitem{Li2008}
Y. Li, V. Bal\'edent, N. Bari\v si\'c, Y. Cho, B. Fauqu\'e, Y. Sidis, G. Yu, X. Zhao, P. Bourges, and M. Greven, 
Nature \textbf{455}, 372 (2008).

\bibitem{Baledent2010}
V. Bal\'edent, B. Fauqu\'e, Y. Sidis, N.~B. Christensen, S. Pailh\`es, K. Conder, E. Pomjakushina, J. Mesot, and P. Bourges, 
Phys. Rev. Lett. \textbf{105}, 027004 (2010).

\bibitem{Almeida2012}
S. De~Almeida-Didry, Y. Sidis, V. Bal\'edent, F. Giovannelli, I. Monot-Laffez, and P. Bourges, 
Phys. Rev. B \textbf{86}, 020504(R) (2012).

\bibitem{Shekhter2013}
A. Shekhter, B.~J. Ramshaw, R. Liang, W.~N. Hardy, D.~A. Bonn, F.~F. Balakirev, R.~D. McDonald, J.~B. Betts, S.~C. Riggs, and A. Migliori, 
Nature \textbf{498}, 75 (2013).

\bibitem{Varma2006}
C.~M. Varma, 
Phys. Rev. B \textbf{73}, 155113 (2006).

\bibitem{Varma2014}
C.~M. Varma, 
J. Phys.: Cond. Matter \textbf{26}, 505701 (2014).

\bibitem{Li2010}
Y. Li \textit{et al.}, 
Nature \textbf{468}, 283 (2010).

\bibitem{Li2012}
Y. Li \textit{et al.}, 
Nature Phys. \textbf{8}, 404 (2012).

\bibitem{Aji2007}
V. Aji and C.~M. Varma, 
Phys. Rev. B \textbf{75}, 224511 (2007).

\bibitem{He2012}
Y. He and C.~M. Varma, 
Phys. Rev. B \textbf{86}, 035124 (2012).

\bibitem{Weber2009}
C. Weber, A. L\"auchli, F. Mila, and Th. Giamarchi, 
Phys. Rev. Lett. \textbf{102}, 017005 (2009).

\bibitem{Mangin2015}
L. Mangin-Thro, Y. Sidis, A. Wildes, and P. Bourges, 
Nature Comm. \textbf{6}, 7705 (2015).

\bibitem{He2011}
R.-H. He \textit{et al.}, 
Science \textbf{331}, 1579 (2011).

\bibitem{Karapetyan2014}
H. Karapetyan, J. Xia, M. H\"ucker, G.~D. Gu, J.~M. Tranquada, M.~M. Fejer, and A. Kapitulnik, 
Phys. Rev. Lett. \textbf{112}, 047003 (2014).

\bibitem{Wang2014}
Y. Wang, A. Chubukov, and R. Nandkishore, 
Phys. Rev. B \textbf{90}, 205130 (2014).

\bibitem{Agterberg2015}
D.~F. Agterberg, D.~S. Melchert, and M.~K. Kashyap, 
Phys. Rev. B \textbf{91}, 054502 (2015).

\bibitem{Mounce2013}
A.~M. Mounce \textit{et al.}, 
Phys. Rev. Lett. \textbf{111}, 187003 (2013).

\bibitem{Strassle2011}
S. Str\"assle, B. Graneli, M. Mali, J. Roos, and H. Keller, 
Phys. Rev. Lett. \textbf{106}, 097003 (2011).

\bibitem{MacDougall2008}
G.~J. MacDougall, A.~A. Aczel, J.~P. Carlo, T. Ito, J. Rodriguez, P.~L. Russo, Y.~J. Uemura, S. Wakimoto, and G.~M. Luke, 
Phys. Rev. Lett. \textbf{101}, 017001 (2008).

\bibitem{Sonier2009}
J.~E. Sonier, V. Pacradouni, S.~A. Sabok-Sayr, W.~N. Hardy, D.~A. Bonn, R. Liang, and H.~A. Mook, 
Phys. Rev. Lett. \textbf{103}, 167002 (2009).

\bibitem{Shekhter2008}
A. Shekhter, L. Shu, V. Aji, D.~E. MacLaughlin, and C.~M. Varma, 
Phys. Rev. Lett. \textbf{101}, 227004 (2008).

\bibitem{Scagnoli2011}
V. Scagnoli, U. Staub, Y. Bodenthin, R.~A. de~Souza, M. Garcia-Fern\'andez, M. Garganourakis, A.~T. Boothroyd, D. Prabhakaran, S.~W. Lovesey, 
Science \textbf{332}, 696 (2011).

\bibitem{Storchak2015}
V.~G. Storchak, J.~H. Brewer, D.~G. Eshchenko, P.~W. Mengyan, O.~E. Parfenov, A.~M. Tokmachev, P. Dosanjh, and S.~N. Barilo, 
Phys. Rev. B \textbf{91}, 205122 (2015).

\bibitem{Nikonov2000}
A.~A. Nikonov and O.~E. Parfenov, 
JETP Letters \textbf{72}, 550 (2000).

\bibitem{Reehuis2006}
M. Reehuis, C. Ulrich, K. Proke\v s, A. Gozar, G. Blumberg, S. Komiya, Y. Ando, P. Pattison, and B. Keimer, 
Phys. Rev. B \textbf{73}, 144513 (2006).

\bibitem{Badoux2016}
S.~Badoux \textit{et al.}, 
Nature \textbf{531}, 210 (2016).

\end{thebibliography}
\end{document}